\documentstyle[letter,psfig,here]{ptptex}

\title{More on Large $Q^2$ Events with Polarized Beams}

\author{Junya {\sc Hashida} and Yuichiro {\sc Kiyo}}

\inst{Deptartment of Physics, Hiroshima University, 
      Higashi-Hiroshima 739-8526, Japan }

\recdate{August 24, 1998
}

\abst{
We discuss polarized proton-positron scattering in the
context of the excess of large $Q^2$ events at HERA.
We define and estimate a polarized asymmetry to examine 
two scenarios, the contact interaction and the stop scenario with 
broken R-parity. This asymmetry exhibits a characteristic behavior depending 
on the scenarios. Thus the polarized experiment at HERA will 
provide with a good test for these models. 
}

\begin{document}

\maketitle

In 1997, an event excess in the neutral current process 
$e^{+}p\rightarrow e^{\prime +} X$ in the region of high 
momentum transfer $Q^2 \geq 15,000$ GeV$^2$ was reported 
by H1 and ZEUS at HERA \cite{H1ZEUS}. The observed cross 
section was $0.71^{+0.14}_{-0.12}$ pb, whereas the standard 
model (SM) predicts $0.49$ pb. The new data \cite{TRAPE} 
analyzed in 1998 are in agreement with the SM up to 
$Q^2\simeq 10,000$ GeV$^2$. The excess at $Q^2 \geq 20,000$ 
GeV$^2$ is not confirmed by the new data but is still present. 
The present situation is rather vague, \cite{TRAPE,ALTA1,VALE} and it 
is still an open question whether this is really an anomalous 
event or if it just results from statistical fluctuation. 
If the excess is not just a statistical fluctuation, 
it must be an indication of new interactions beyond the SM, 
because it appears to be very difficult to explain the data in the 
framework of the SM. 

There have appeared many proposals and analyses of this problem.
New contact interactions (CI) stemming from high energy scale physics
have been analyzed, \cite{ALTA,BARG,DESH} and supersymmetric (SUSY) models 
with R-parity violating ($R_{p}\hspace{-11pt}/~~$) interactions have 
also been discussed. \cite{ALTA} 
The two-stop scenario, \cite{KON} left stop $\tilde{t}_{L}$ 
is a mixture of the almost degenerate mass eigenstates 
of $\tilde{t}_{1}$ and $\tilde{t}_{2}$, with $R_{p}\hspace{-11pt}/~~$ 
interactions was proposed as one of the candidates to explain broad mass 
distribution in the data. 

HERA will begin a polarized experiment, \cite{EXP} polarized proton 
$p(\uparrow/\downarrow)$ and lepton (positron in our discussion) 
scattering, in the near future.
The polarized experiment is important because the polarization 
of the proton and lepton beams make it possible to test the 
chiral structure of the interactions. \cite{VIRE}
Thus it is interesting to ask what HERA will teach us about the 
models in the future polarized program. 

In this paper, we examine two scenarios, the CI and the two-stop scenarios 
in the context of the large $Q^{2}$ events at the polarized HERA.
Our interest is in determining how we can examine these scenarios and what 
the characteristics of the models are. 
Thus we discuss these scenarios with regard to the future polarized experiment 
$e^{+} p(\uparrow/\downarrow) \rightarrow e^{+ \prime}X$.   
After giving the model Lagrangians, we calculate the parton level 
cross sections which will be convoluted with parton distributions
to form the physical cross section. 

The Lagrangian for the CI \cite{ALTA,BARG,DESH} assumes the form
\begin{eqnarray}
L_{CI}
&=&
\frac{4 \pi}{\Lambda^2}
\sum_{\stackrel{q=u,d}{a,b=L,R}}
\eta^q_{ab}
\left( 
\bar{e}_{a}\gamma^\mu e_{a}
\right)
\left(
\bar{q}_{b}\gamma_\mu q_{b}
\right),
\end{eqnarray}
which is the effective interaction of a certain underlying high energy physics
describing low energy phenomena in the neutral current process. 
The subscript $R (L)$ denotes the chirality of the fields, $\eta^q_{ab}=\pm 1, 0$, 
and $\Lambda$ is the  mass scale of a heavy particle which might be exchanged among 
quarks and leptons. Thus these interactions are suppressed by the mass scale 
of the new physics, and some constraints \cite{TRAPE,GCHO} have been obtained 
for $\Lambda$ in many experiments. 
The superpotential, for the stop scenario with $R_{p}\hspace{-11pt}/~~$ 
interaction \cite{ALTA,KON}, is
given by
\begin{equation}
W_{R\hspace{-6pt}/~}
=
\lambda^{\prime}_{131}
L_{1} Q_{3} D^{c}_{1},
\end{equation}
where $L_{1}$ and $Q_{3}$ are the superfields of the $SU(2)_L$ lepton and 
quark doublet, respectively, and $D^{c}_{1}$ is the singlet down type quark. 
Here the subscripts 1, 2 and 3 are the generation indices. 
The interaction Lagrangian can be obtained from the superpotential
\begin{equation}
L_{\lambda^\prime}
=
\lambda^{\prime}_{131}
\left(
\tilde{t}_{L} \bar{d}P_{L}e + 
\tilde{e}_{L} \bar{d}P_{L}t + 
\bar{\tilde{d}}_{R}\bar{e}^{c}P_{L}t 
-
\tilde{b}_{L}\bar{d}P_{L}\nu_{e}-
\tilde{\nu}_{L}\bar{d}P_{L}b -
\bar{\tilde{d}}_{R}\bar{\nu}^{c}_{e} P_{L}b
\right) 
+ h.c.  
\end{equation}
For the scalar fields, $R (L)$ denotes the chirality of their superpartners.
We discuss the proton-positron scattering, so only the first term 
$\tilde{t}_{L} \bar{d}P_{L}e + h.c.$ is relevant. 
In the two-stop scenario, 
the left stop $\tilde{t}_L$ is the superposition of the two mass eigenstates 
$\tilde{t}_1$ and $\tilde{t}_2$ with the mixing angle $\theta_t$; namely 
$\tilde{t}_L= \tilde{t}_1 \cos\theta_t - \tilde{t}_2 \sin\theta_t$.
The stop $\tilde{t}_{L}$ can couple only to the left handed lepton field 
$e_{L}$ and the right handed down quark $d_{R}$.
This is an important point in our discussion, because the polarized 
experiment can distinguish the chiral structure of the interactions
in the parton-lepton scattering. 

The partonic cross sections $\hat{\sigma}$ for the models are given by
\begin{eqnarray}
\frac{ d \hat{\sigma}(e^{+}_{I} f_{J}) }{dx_B dQ^{2}}
&=&
\delta(x_B - x)\frac{(4\pi \alpha_{e})^{2}}{8 \pi \left( \hat{s} ~ Q^{2} \right)^{2} }
\left[
(1+ I \cdot J )\hat{s}^{2} + (1- I\cdot J) \hat{u}^{2}
\right]
\nonumber \\
&\times&
\left| 
Q_{\gamma}(e)Q_{\gamma}(f) 
+ \frac{Q_{Z}^{-I}(e)Q_{Z}^{J}(f)
       }{\sin^2\theta_W}
\frac{Q^{2}}{Q^{2}+M_{Z}^{2}}
+\Delta
\right|^{2},
\end{eqnarray}
where $I(J)=\pm$ correspond to the helicities $\pm 1/2$ of the 
positron (quark), $x_B$ is the Bjorken variable and 
$x$ is the momentum fraction of the parton,
$\alpha_{e}=e^{2}/(4\pi)$, $\theta_W$ is the electro-weak angle, 
and $\hat{s}$ and $\hat{u}$ are the Mandelstam variables with respect to the 
parton-positron system, which are defined by $\hat{s}=x s$ and $\hat{u}=x u$.
$\Delta$ is the contribution from the CI or $R_{p}\hspace{-11pt}/~~$ 
interaction.
We neglect the masses of the quarks and positron in this paper.
The coupling constants of the electron and  up and down quarks to the photon and Z boson 
are given by
\begin{eqnarray}
Q_{\gamma}(e)&=&-1,~~~
Q_{Z}^{+}(e)=
\frac{\sin^2\theta_W}{\cos^2\theta_W},~~~
Q_{Z}^{-}(e)=
\frac{2 \sin^2\theta_W-1}{2 \cos\theta_W},~~~
\\
Q_{\gamma}(u)&=&\frac{2}{3},~~~
Q_{Z}^{+}(u)=
\frac{-2 \sin^2\theta_W}{3 \cos\theta_W},~~~
Q_{Z}^{-}(u)=
\frac{3- 4 \sin^2\theta_W}{6 \cos\theta_W},~~~
\\
Q_{\gamma}(d)&=&\frac{-1}{3},~~~
Q_{Z}^{+}(d)=
\frac{\sin^2\theta_W}{3 \cos^2\theta_W},~~~
Q_{Z}^{-}(d)=
\frac{-3+ 2 \sin^2\theta_W}{6 \cos\theta_W}.
\end{eqnarray} 
For the CI scenario, $\Delta$ is given by 
\begin{equation}
\Delta(Q^2)=-\frac{Q^2 \eta^q_{-IJ} }{\alpha_e \Lambda^2},
\end{equation}
where the subscripts $+$ and $-$ of the $\eta^q_{ab}$ correspond
respectively to $R$ and $L$. 
The stop exchange with the $R_{p}\hspace{-11pt}/~~$ interaction yields
the following contribution
\begin{equation}
\Delta(\hat{s},Q^2)
=
- \frac{ \alpha_{131} Q^2
       }{2 \alpha_{e}}
\left( \frac{\cos^{2}\theta_{t}
            }{\hat{s}-\tilde{m}_{1}^{2}+i\tilde{m}_{1}\Gamma_{\tilde{t}_{1}}}
      	+
       \frac{\sin^{2}\theta_{t}
            }{\hat{s}-\tilde{m}_{2}^{2}+i\tilde{m}_{2}\Gamma_{\tilde{t}_{2}}} 
\right)
\end{equation}
for the $I=J=+$ channel and $f=d$. Otherwise $\Delta=0$ in the stop scenario. 
Here $\alpha_{131}=\lambda^{\prime 2}_{131}/(4\pi)$,
$\tilde{m}_{1,2}$ and $\Gamma_{ \tilde{t}_{1,2}}$ are the masses and 
widths of $\tilde{t}_{1,2}$ respectively. 

The cross section for the polarized proton-positron scattering is
obtained by convoluting the partonic cross sections with the polarized 
parton distribution functions. 
The cross section $\sigma(e^{+}p(\uparrow))$ for the longitudinally 
polarized proton $p(\uparrow)$ and positron scattering can be written:
\begin{equation}
\frac{ d \sigma }{dx_B dQ^{2}}(e^{+}p(\uparrow) ) 
=
\int^{1}_{0} d x
\sum_{f}
\left(
\frac{d\hat{\sigma} \left( e^{+} f_{+} \right) 
      }{dx_B dQ^{2}}
f_{+/\uparrow}(x)
+
\frac{d\hat{\sigma} \left( e^{+} f_{-} \right) 
      }{dx_B dQ^{2}}(x)
f_{-/\uparrow}
\right),
\end{equation}
where $f_{\pm/\uparrow}(x)$ is the polarized parton distribution 
function for the flavor $f$ parton with momentum fraction $x$ 
and helicity $\pm1/2$ in the proton $p(\uparrow)$.

We are interested in the region which is characterized by 
the two variables $Q^{2}$ and the invariant mass $M = \sqrt{\hat{s}}$, 
with $Q^{2} \geq 20,000$ GeV$^2$ and $M \sim 200$ GeV. 
This corresponds to the region in which the partons in 
the proton have a momentum fraction $x \sim 0.4$.
Thus we can safely neglect the contribution from the sea quarks, 
because their distribution is quite small in that region, and 
contributions from gluons are next to leading order in the QCD coupling
constant.

\begin{figure}[t]
\begin{center}
\leavevmode\psfig{file=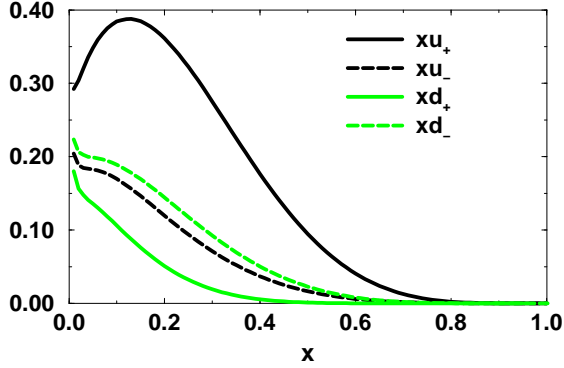,width=10cm,angle=-90}
\caption{ The polarized parton distributions
$x u_{\pm/\uparrow}(x)$ and $x d_{\pm/\uparrow}(x)$ 
at the scale $Q^{2}=20,000$ GeV$^2$.}
\label{fig:pd}
\end{center}
\end{figure}

In Fig.\ref{fig:pd}, we show the polarized parton distributions
$x u_{\pm/\uparrow}(x),x d_{\pm/\uparrow}(x)$ in the proton as a 
function of the momentum fraction $x$. 
The scale of the distributions is taken to be $Q^{2}=20,000$ GeV$^2$ 
using the parameterization of Refs. \cite{GS} and \cite{MRS}. 
Our numerical estimation has shown that the effects of $Q^2$ evolutions 
to the parton distributions were tiny in the region considered 
in this paper. This is reasonable because the 
change in the QCD coupling constant is small for large $Q^2$.

One can see that most of the down quarks are oppositely 
polarized ($x d_{+}(x) \leq x d_{-}(x) $) with respect to the proton spin, 
while the up quarks are polarized along the proton spin 
($x u_{-}(x) \leq x u_{+}(x) $). 
%
The largest component of the proton in the region of interest 
is the up quark with helicity $+1/2$. 
In the CI scenario, the up quarks might contribute to the large $Q^2$ 
excess if $\eta_{ab}^u$ is sufficiently large. 
However, in the stop scenario, they can not contribute, because there is no 
coupling between the up quark and stop.
The next large component is the down quark with helicity 
$-1/2$. This also does not couple to the stop. The situation changes 
if we use an oppositely polarized proton beam $p(\downarrow)$, 
because in this case the down quarks with helicity $+1/2$ represent the next 
largest component in the proton $p(\downarrow)$.
Hence the cross section for $e^{+} p(\downarrow)\rightarrow e^{+\prime}X$
is larger than that for $e^{+} p(\uparrow)\rightarrow e^{+\prime}X$
in the stop scenario.

\begin{figure}[H]
\begin{center}
\leavevmode\psfig{file=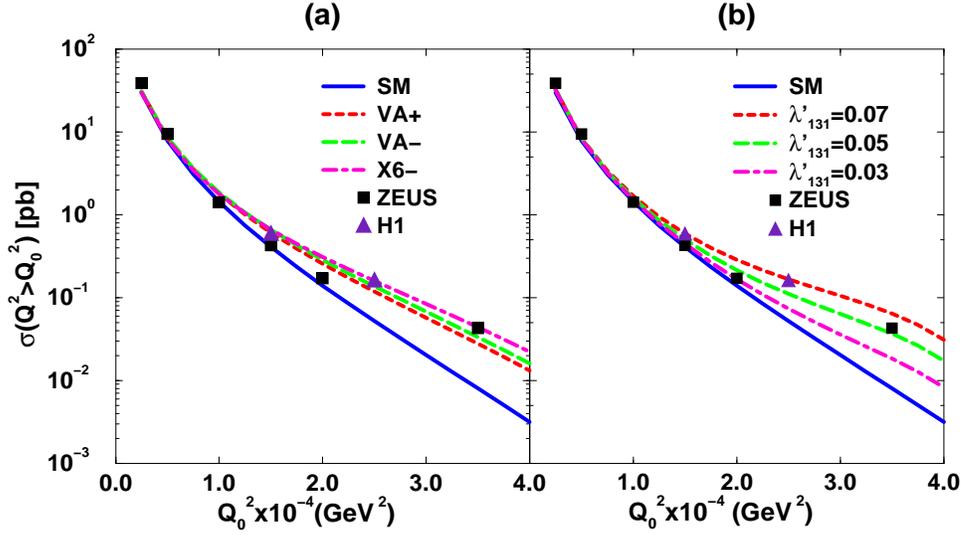,width=13cm,angle=-90}
\caption{The unpolarized cross sections in the range $ 2,500 \leq Q^2_0 \leq 40,000$ GeV$^2$
for (a) the CI scenario with parameter sets VA+, VA-, X6-, and 
(b) the two-stop scenario with 
$\lambda^\prime_{131}=0.07,0.05,0.03$. 
The solid lines indicate the cross sections predicted by the SM. }
\label{fig:unpol}
\end{center}
\end{figure}

To begin with, we give numerical results for the unpolarized case.
Figure \ref{fig:unpol} shows how the CI and stop scenarios explain 
the unpolarized HERA data. \cite{TRAPE} The following three parameter 
sets for the CI scenario have been employed:
\begin{center}
$
(1) ~~
VA+ :
(\eta^{u/d}_{LL},\eta^{u/d}_{LR},\eta^{u/d}_{RL},\eta^{u/d}_{RR})
=(+,-,+,-)
$ 
~~with $\Lambda=2.8$ TeV, \\
$
(2) ~~
VA- :
(\eta^{u/d}_{LL},\eta^{u/d}_{LR},\eta^{u/d}_{RL},\eta^{u/d}_{RR})
=(-,+,-,+)
$
~~with $\Lambda=2.8$ TeV,\\
$
(3) ~~~
X6- :
(\eta^{u/d}_{LL},\eta^{u/d}_{LR},\eta^{u/d}_{RL},\eta^{u/d}_{RR})
=(0,0,-,+)
$ 
~~~~with $\Lambda=1.9$ TeV,\\
\end{center}
where the 95\% confidence level limits \cite{TRAPE} on $\Lambda$ obtained by 
the ZEUS and H1 collaborations were used. The value of $\Lambda$ for the other 
parameter sets are very strongly constrained by the results of other experiments. 
\cite{GCHO}
For the two-stop scenario, we used the values 
$\lambda_{131}^\prime=0.07, 0.05, 0.03$, 
$\tilde{m}_{1,2}=200, 230$ GeV, and the stop mixing angle 
$\cos\theta_{t}=0.5$. The branching ratios 
$Br(\tilde{t}_{1,2} \rightarrow e^+ d )$ are 0.65 and 1.0 for 
$\tilde{t}_{1}$ and $\tilde{t}_{2}$, respectively. \cite{KON,ASAK} 
The parameter sets of the CI and the stop scenarios
account for the unpolarized experimental data quite well.
It is also seen that the behavior of the unplarized cross sections 
are similar in both the scenarios.

Next, we discuss the polarized case. 
For our purposes, it is useful to introduce the spin asymmetry 
$\cal{A}$, which is defined by 
\begin{eqnarray}
{\cal A}(Q_{0}^2) 
&\equiv&
\frac{\int \sigma_{\uparrow}-\int \sigma_{\downarrow}
    }{\int \sigma_{\uparrow}+\int \sigma_{\downarrow}},
\end{eqnarray}
where $\int \sigma_{\uparrow/\downarrow}$ is the integrated cross section
for the polarized proton $p(\uparrow/\downarrow)$ and positron
scattering. We have 
\begin{eqnarray}
\int \sigma_{\uparrow/\downarrow}
&=&
\int_{Q_{0}^2}^{Q_{max}^2} dQ^2 dx_{B}
~\sigma(p(\uparrow/\downarrow)e^+\rightarrow e^{\prime +}X),
\end{eqnarray}
where $Q_{max}^{2}=90,000$ GeV$^2$.
%
%
\begin{figure}[H]
\begin{center}
\leavevmode\psfig{file=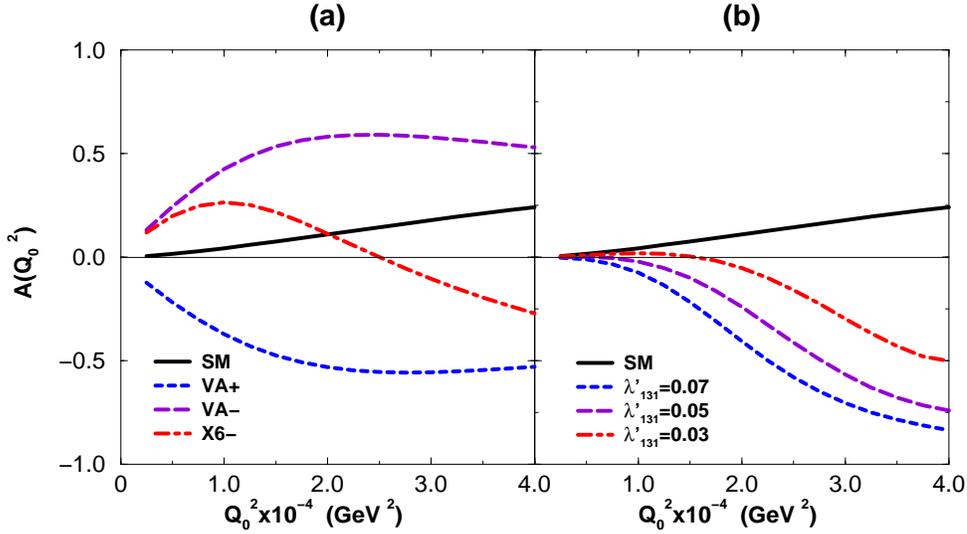,width=13cm,angle=-90}
\caption{ 
The asymmetries $ {\cal A}(Q_0^2)$ in the range $ 2,500 \leq Q^2_0 \leq 40,000$ GeV$^2$
for (a) the CI scenario with parameter sets VA+, VA-, X6-, 
and (b) the two-stop scenario with $\lambda^\prime_{131}=0.07, 0.05, 0.03$. 
The solid lines correspond to the prediction of the SM. 
}
\label{fig:asy}
\end{center}
\end{figure}
%
In Fig.\ref{fig:asy}, we plot the spin asymmetry for the CI and two-stop
scenarios. One can see that $\cal A$ in the stop scenario has a large negative 
value in the large $Q_0^2$ region, while the SM prediction is positive in 
the region $Q^2_0 \geq 2,500 $ GeV$^2$.  
This is because the proton $p(\uparrow)$ contains more down quarks with helicity $-1/2$ 
than with helicity $+1/2$ and only the down quark with helicity 
$+1/2$ can couple to the stops which produce the large contribution to the cross 
sections. This is a characteristic feature for the stop scenario with
the $R_{p}\hspace{-11pt}/~~$ interaction. A different choice for the parameters in 
the stop scenario does not change the results appreciably.   
The asymmetry for the parameter set VA+ in the CI scenario has a negative 
value and is similar to that in the stop scenario. 
However, the asymmetry for VA+ has negative value even when  
$Q^2_0 \sim 2,500$ GeV$^2$, where the asymmetry in the stop scenario 
is nearly zero. The asymmetries for the CI scenario are very different 
from that for the SM, even at $Q^2_0 \sim 2,500$ GeV$^2$.  
Hence, observing these behavior, the two scenarios may be considered distinguishable.

In summary, we discussed polarized proton-positron scattering in 
the context of the excess of large $Q^2$ events at HERA.
For the CI scenario, the asymmetries $\cal{A}$ exhibit distinctive behavior 
for $Q^2_0\sim 2,500$ GeV$^2$, and the value depends on the parameter sets. 
For the two-stop scenario, there is a characteristic dependence on 
$Q_0^2$: the value changes from zero to near -1 as $Q_0^2$ becomes 
larger. Studying this behavior at future polarized HERA will provide a good test for 
these models. 

The authors thank to T. Nasuno, H. Tochimura and Y. Yasui for 
helpful discussions, and to J. Kodaira for useful comments and 
reading manuscript.   

%

\end{document}